\documentclass[english]{revtex4-1}
\usepackage[T1]{fontenc}
\usepackage[latin9]{inputenc}
\usepackage{float}
\usepackage{amsmath}
\usepackage{amssymb}
\usepackage{graphicx}

\makeatletter
\usepackage{babel}

\makeatother

\usepackage{babel}
\begin{document}

\title{Measurability of Coulomb wavepacket scattering effects}

\author{Scott E. Hoffmann}

\address{School of Mathematics and Physics,~~\\
 The University of Queensland,~~\\
 Brisbane, QLD 4072~~\\
 Australia}
\email{scott.hoffmann@uqconnect.edu.au}

\begin{abstract}
A previous paper {[}J. Phys. B: At. Mol. Opt. Phys. \textbf{50}, 215302
(2017){]} showed that partial wave analysis becomes applicable to
nonrelativistic Coulomb scattering if wavepackets are used. The scattering
geometry considered was special: that of a head-on collision between
the wavepacket and the centre of the potential. Our results predicted,
in this case, a shadow zone of low probability for small angles around
the forward direction for the description of alpha scattering from
a gold foil. In this paper we generalize the results to the case of
a nonzero impact parameter, a displacement of the wavepacket centre
perpendicular to the average momentum direction. We predict a large
flux in the forward direction from events with large impact parameters.
We find a significant probability of scattering into the deviation
region for impact parameters of order the spatial width of the wavepacket.
Averaging over impact parameters produces predictions in excellent
agreement with the Rutherford formula down to lower angles than for
the zero impact parameter prediction. We consider issues that would
arise in a real experiment and discuss the possibility of measuring
a deviation from the Rutherford formula.
\end{abstract}
\maketitle

\section{Introduction}

In a previous paper \cite{Hoffmann2017a}, this author developed the
scattering theory for a wavepacket in a Coulomb potential. The method
used was partial wave analysis, which had been thought to be inapplicable
to the Coulomb potential because the sum over the angular momentum
index, $l,$ diverges for a plane wave treatment. The use of wavepackets
was found to introduce a convergence factor into that sum. Then we
summed the series numerically for a variety of cases, using the phase
shifts obtained from the exact solution for the partial wave energy
eigenvectors \cite{Messiah1961}.

We found generally excellent agreement with the Rutherford scattering
cross section,
\begin{equation}
\frac{d\sigma}{d\Omega}_{\mathrm{Rutherford}}=\frac{Z_{1}^{2}Z_{2}^{2}\alpha^{2}}{16E^{2}\sin^{4}(\frac{\theta}{2})},\label{eq:1.1}
\end{equation}
but found deviations in all cases, at low scattering angles. Here
$Z_{1}$ and $Z_{2}$ are the atomic numbers of the target and projectile,
respectively, $\alpha$ is the fine structure constant, $E$ is the
incident energy and $\theta$ is the scattering angle.

Our method calculates probabilities of wavepacket-to-wavepacket transitions.
We derived a formula that relates these probabilities to the differential
cross section (see Eq. (\ref{eq:2.24})). Since a probability is constrained
to be less than unity while the Rutherford formula diverges in the
forward direction, a disagreement was inevitable.

The aim of this paper is to investigate these disagreements with the
Rutherford formula to see if any of them might be experimentally measurable.

\begin{figure}[H]
\begin{centering}
\includegraphics[width=7cm]{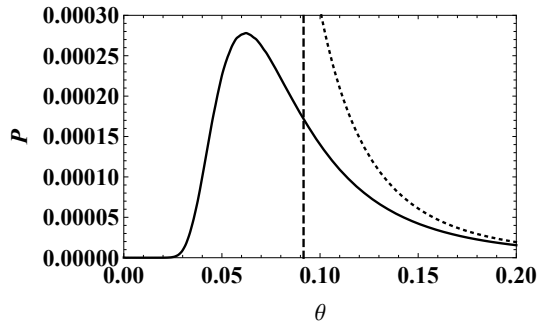}
\par\end{centering}
\caption{\label{fig:Shadow-zone-for}Shadow zone for $E=4.8\,\mathrm{MeV},$
alpha on gold, $\epsilon=0.001.$ Probability (solid) compared to
the Rutherford probability (dotted). The vertical, dashed, line shows
our estimate, $\theta_{D}=4\epsilon|\eta|,$ for the size of the region
of deviation from the Rutherford formula. This is for a head-on collision
(zero impact parameter).}

\end{figure}

For incident energies typical of scattering experiments that have
been performed, we predict small probabilities of scattering into
a small angular region around the forward direction, which we call
a shadow zone. This is shown in Figure \ref{fig:Shadow-zone-for}
for alpha particles of incident energy $E=4.8\,\mathrm{MeV}$ on gold
foil with a momentum resolution parameter $\epsilon=0.001$ (see Eq.
(\ref{eq:2.5})).

This prediction is in disagreement with the experimental observation
of a high particle flux peaked around the forward direction. In this
paper, we argue that the reason for this disagreement is the special
scattering geometry that was chosen for the previous calculations.
We modelled only a head-on collision, with the centre of the wavepacket
approaching the centre of the potential. In this paper we investigate
the effects of nonzero impact parameters, displacements perpendicular
to the average direction of motion. Our physical expectation is that
for an event with a sufficiently large impact parameter, the wavepacket
will pass the potential largely undisturbed to contribute to a strong
peak around the forward direction. The shadow zone around the forward
direction for zero impact parameter is then seen as an interesting
feature of the wave nature of the scattering, but may not be experimentally
observable.

To proceed with these investigations, in Section \ref{sec:Wavefunctions-for-nonzero},
we construct the wavefunctions and then the scattering probability
for nonzero impact parameters. In Section \ref{sec:Scattering-into-the},
we use this probability to confirm the result just postulated, that
at sufficiently high impact parameter, the wavepacket emerges largely
undisturbed, with nearly unit probability, moving in the forward direction.

In Section \ref{sec:Behaviour-at-small}, we consider a model Coulomb
scattering experiment, very similar to the original experiments of
Geiger, Marsden and Rutherford \cite{Geiger1909,Rutherford1911}.
It is only necessary to have better collimation of the beam of alpha
particles to resolve details at low angles. To model an experiment,
it is necessary to integrate probabilities over impact parameters.
This is done for three angles in the deviation region and the results
compared to the Rutherford prediction.

Conclusions follow in Section \ref{sec:Conclusions}.

\section{\label{sec:Wavefunctions-for-nonzero}Scattering probability and
cross section for nonzero impact parameter}

\begin{figure}
\begin{centering}
\includegraphics[width=15cm]{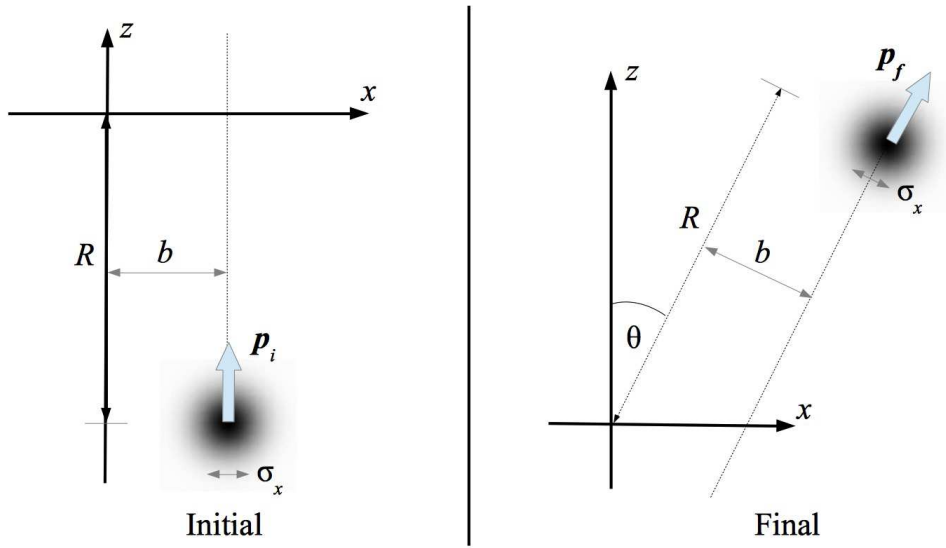}
\par\end{centering}
\caption{\label{fig:Scattering-geometry-with}Scattering geometry with nonzero
impact parameters.}

\end{figure}

The geometry we consider, modified from \cite{Hoffmann2017a} to include
nonzero impact parameters, is shown in Figure \ref{fig:Scattering-geometry-with}.

Our method involves first transforming the free wavefunction of the
incident projectile from a basis of eigenvectors of momentum, $\boldsymbol{k},$
to a basis of eigenvector of momentum magnitude, $k,$ and angular
momentum, with indices $l,m.$ The transformation formula found in
the earlier paper \cite{Hoffmann2017a} is
\begin{equation}
\Psi_{\mathrm{free}}(k,l,m)=k\int_{0}^{\pi}\sin\theta_{k}d\theta_{k}\int_{0}^{2\pi}d\varphi_{k}\,Y_{lm}^{*}(\theta_{k},\varphi_{k})\psi_{\mathrm{free}}(\boldsymbol{k}).\label{eq:2.1}
\end{equation}
The spherical harmonic can be written in terms of a Wigner rotation
matrix
\begin{equation}
Y_{lm}^{*}(\theta_{k},\varphi_{k})=\sqrt{\frac{2l+1}{4\pi}}e^{-im\varphi}d_{m0}^{l}(\theta).\label{eq:2.2}
\end{equation}
Choosing propagation in the $z$ direction with an impact parameter
$\boldsymbol{b}=b\,\hat{\boldsymbol{x}}$ perpendicular to $\hat{\boldsymbol{z}}$
gives the simplest results. The $\boldsymbol{k}$ wavefunction for
this geometry is chosen as
\begin{equation}
\psi_{\mathrm{free}}(\boldsymbol{k})=e^{-i\boldsymbol{k}\cdot\boldsymbol{b}}e^{-i\boldsymbol{k}\cdot\boldsymbol{R}_{i}}\frac{e^{-|\boldsymbol{k}-\boldsymbol{p}_{i}|^{2}/4\sigma_{p}^{2}}}{(2\pi\sigma_{p}^{2})^{\frac{3}{4}}}\label{eq:2.3}
\end{equation}
with
\begin{equation}
\boldsymbol{p}_{i}=p\,\hat{\boldsymbol{z}},\quad\boldsymbol{R}_{i}=-R\,\hat{\boldsymbol{z}}.\label{eq:2.4}
\end{equation}
The standard deviation of momentum is $\sigma_{p}$ in all directions.
We choose the momentum resolution parameter
\begin{equation}
\epsilon=\frac{\sigma_{p}}{p}\label{eq:2.5}
\end{equation}
to be much less than unity. The corresponding position wavefunction
at $t=0$ has width $\sigma_{x}$ in all directions, with $\sigma_{x}\sigma_{p}=1/2.$
With the particular choice
\begin{equation}
R=\frac{1}{\sqrt{\epsilon}}\,\sigma_{x},\label{eq:2.6}
\end{equation}
we ensure that the initial wavepacket is far from the origin (compared
to the width), growing farther as $\epsilon$ is made smaller, but
wavepacket spreading remains negligible over the course of the scattering
experiment. Then $\sqrt{\epsilon}$ becomes the small parameter for
our approximations.

In order to apply approximations, we impose a limit on the impact
parameter
\begin{equation}
b\leq\frac{1}{\sqrt{\epsilon}}\,\sigma_{x}.\label{eq:2.7}
\end{equation}
Then we expand
\begin{equation}
\boldsymbol{k}\cdot\boldsymbol{b}=bp\,\theta_{k}\cos\varphi_{k}+\mathcal{O}(\sqrt{\epsilon}).\label{eq:2.8}
\end{equation}
Then the integral over $\varphi_{k}$ evaluates to (\cite{Gradsteyn1980},
their Eq. (8.411.1))
\begin{equation}
\int_{0}^{2\pi}d\varphi_{k}\,e^{-im\varphi_{k}}e^{-ibp\,\theta_{k}\cos\varphi_{k}}=2\pi\,e^{-i|m|\pi/2}J_{|m|}(bp\,\theta_{k}).\label{eq:2.9}
\end{equation}

Next we use the low-angle approximation of the Wigner rotation matrix,
uniform in $l,m,$ derived in \cite{Hoffmann2018b},
\begin{equation}
d_{m0}^{l}(\theta_{k})\sim\Phi(m)\,[\frac{(l+|m|)!}{(l-|m|)!}]^{\frac{1}{2}}\frac{1}{\Lambda^{|m|}}J_{|m|}(\Lambda\theta_{k}),\label{eq:2.10}
\end{equation}
with
\begin{equation}
\Lambda(l,m)=\sqrt{(l+\frac{1}{2})^{2}-\frac{1}{3}m^{2}+\frac{1}{12}}\label{eq:2.11}
\end{equation}
and
\begin{equation}
\Phi(m)=\begin{cases}
(-)^{m} & m\geq0,\\
1 & m<0.
\end{cases}\label{eq:2.12}
\end{equation}
For $m=0$ and $l=2000,$ the absolute error in this approximation
is less than $5\times10^{-10}$ on $0\leq\theta\leq0.2.$

We use the integral (\cite{Gradsteyn1980}, their Eq. (6.633.2))
\begin{equation}
\int_{0}^{\pi}\theta_{k}\,d\theta_{k}\,e^{-p^{2}\theta_{k}^{2}/4\sigma_{p}^{2}}\,J_{|m|}(\Lambda\theta_{k})J_{|m|}(bp\,\theta_{k})=2\epsilon^{2}e^{-\epsilon^{2}(\Lambda-bp)^{2}}e^{-2\epsilon^{2}\Lambda bp}I_{|m|}(2\epsilon^{2}\Lambda bp).\label{eq:2.13}
\end{equation}
Noting
\begin{equation}
\Phi(m)\,e^{-i|m|\pi/2}=e^{im\pi/2},\label{eq:2.14}
\end{equation}
we find the free wavefunction in the $k,l,m$ basis
\begin{equation}
\Psi_{\mathrm{free}}(k,l,m)=e^{im\pi/2}e^{+ikR}\frac{e^{-(k-p)^{2}/4\sigma_{p}^{2}}}{(2\pi\sigma_{p}^{2})^{\frac{1}{4}}}2\epsilon\sqrt{l+\frac{1}{2}}[\frac{(l+|m|)!}{(l-|m|)!}]^{\frac{1}{2}}\frac{1}{\Lambda^{|m|}}e^{-\epsilon^{2}(\Lambda-bp)^{2}}\mu_{|m|}(2\epsilon^{2}\Lambda bp),\label{eq:2.15}
\end{equation}
with
\begin{equation}
\mu_{|m|}(z)\equiv e^{-z}\,I_{|m|}(z).\label{eq:2.16}
\end{equation}

The following steps are very similar to those in \cite{Hoffmann2017a},
so will not be repeated here. Only phase shifts can be applied to
the free $k,l,m$ wavefunction to produce the incoming $k,l,m$ wavefunction,
to preserve
\begin{equation}
|\Psi_{\mathrm{in}}(k,l,m)|^{2}\rightarrow|\Psi_{\mathrm{free}}(k,l,m)|^{2}\quad\mathrm{as}\ \epsilon\rightarrow0^{+}.\label{eq:2.17}
\end{equation}
Those phase shifts are found in terms of the Coulomb phase shifts
\cite{Messiah1961}
\begin{equation}
e^{i2\sigma_{l}(k)}=\frac{\Gamma(l+1+i\eta(k))}{\Gamma(l+1-i\eta(k))}\label{eq:2.18}
\end{equation}
(with $\eta(k)=\alpha/(k/m_{0})$ a dimensionless measure of the strength
of the interaction) and the expansion of the logarithmic term, $\eta(k)\ln(2kr),$
that appears in the asymptotic approximation of the Coulomb spherical
waves. Applying this phase shift ensures that the position probability
density of the interacting state vector satisfies
\begin{equation}
|\psi_{\mathrm{in}}(\boldsymbol{r})|^{2}\rightarrow|\psi_{\mathrm{free}}(\boldsymbol{r})|^{2}\quad\mathrm{as}\ \epsilon\rightarrow0^{+}.\label{eq:2.19}
\end{equation}
The outgoing state vector is constructed from the incoming by applying
the antiunitary time reversal operator (which complex conjugates the
phase shifts) and a rotation into the final scattering angle, $\theta.$
Note that by conservation of angular momentum, the final state will
have the same average impact parameter as the initial state.

We find the result for the probability of a wavepacket to wavepacket
transition
\begin{equation}
P(\theta,\eta,\beta,\delta,\epsilon)=\left|\sum_{l=0}^{\infty}\sum_{m_{f}=-l}^{l}\sum_{m_{i}=-l}^{l}\Phi_{\mathrm{free}}(l,m_{f})e^{+i(m_{i}-m_{f})\pi/2}d_{m_{i}m_{f}}^{(l)}(\theta)\Phi_{\mathrm{free}}(l,m_{i})e^{i2\sigma_{l}(p)}e^{-(\delta-\xi_{l})^{2}/8}\right|^{2},\label{eq:2.20}
\end{equation}
where
\begin{equation}
\xi_{l}\equiv4\epsilon\eta(p)\{\ln(2pR)-1-\frac{\partial\sigma_{l}[\eta(p)]}{\partial\eta}\}\label{eq:2.21}
\end{equation}
and
\begin{align}
\Phi_{\mathrm{free}}(l,m) & =2\epsilon\sqrt{l+\frac{1}{2}}\,[\frac{(l+|m|)!}{(l-|m|)!}]^{\frac{1}{2}}\frac{1}{\Lambda^{|m|}}e^{-\epsilon^{2}(\Lambda-bp)^{2}}\mu_{|m|}(2\epsilon^{2}\Lambda bp)\label{eq:2.22}
\end{align}
and $\beta=b/\sigma_{x}$ is a dimensionless measure of the impact
parameter. Here
\begin{equation}
\delta(T)=\frac{\frac{p}{m_{0}}T-2R}{\sigma_{x}}\label{eq:2.23}
\end{equation}
in terms of the interaction time, $T,$ and the mass, $m_{0},$ of
the projectile. Plotting the probability versus $\delta$ shows a
peak shifted in time, with the position of the maximum, $\delta_{\mathrm{max}},$
showing a time delay if $\delta_{\mathrm{max}}>0$ and an advancement
if $\delta_{\mathrm{max}}<0.$ In practice, we find the value, $\delta_{\mathrm{max}},$
that maximizes the probability, then recalculate the probability with
$\delta$ set to that value.

According to a formula derived in \cite{Hoffmann2017a}, the differential
cross section is related to the probability by
\begin{equation}
\frac{d\sigma}{d\Omega}=\frac{p^{2}}{16\,\sigma_{p}^{4}}P(\theta,\eta,\beta,\delta,\epsilon),\label{eq:2.24}
\end{equation}
valid for Gaussian wavepackets. We apply this to the Rutherford differential
cross section to form a dimensionless ``probability'', not a true
probability since it rises greater than unity,
\begin{equation}
P_{\mathrm{Ruth}}(\theta,\eta,\epsilon)=\frac{4\epsilon^{4}\eta^{2}}{\sin^{4}\frac{\theta}{2}}.\label{eq:2.25}
\end{equation}

\section{\label{sec:Scattering-into-the}Scattering into the forward direction}

We start by considering the question of the influence of nonzero impact
parameters on the scattering into the forward direction.

We consider the original experiment performed by Geiger, Marsden and
Rutherford \cite{Geiger1909,Rutherford1911}, with an alpha particle
of energy $E=4.8\,\mathrm{MeV}$ from Radium-226 incident on a thin
gold foil. This gives the strength parameter $\eta=22.8.$ (Note this
was incorrectly stated as $\eta=23.1$ in \cite{Hoffmann2017a}.)
We choose the momentum resolution parameter $\epsilon=0.001$ which
gives $\sqrt{\epsilon}=0.032.$ Note that energy linewidths as small
as $\Delta E=2\,\mathrm{keV}$ have been observed \cite{Pomme2015}
using extremely thin radium samples. This would imply that the momentum
resolution parameter was $\epsilon\leq2.1\times10^{-4}$ for those
sources.

The alpha particle wavepackets (of spatial width $\sigma_{x}=0.0052\,\textrm{Å}$)
will approach the gold nuclei with every impact parameter between
0 and approximately half the internuclear distance of $2.6\,\textrm{Å}$
(corresponding to $\beta=250$ with the parameters we have chosen).
Note that the optimal starting separation from the nucleus is only
$R=0.16\,\textrm{Å},$ indicating that it is very difficult, in practice,
to create an ideal scattering experiment where wavepacket spreading
is negligible.

For those events with small impact parameters, we expect that there
will be scattering into all directions. In an event with sufficiently
large impact parameter, it is our physical hypothesis that the wavepacket
will pass the nucleus largely undisturbed and contribute to a peak
of flux in the forward direction.

We confirm this expectation now by calculating the probabilities for
a range of impact parameters of scattering into the forward direction.
In that case
\begin{equation}
d_{m_{i}m_{f}}^{(l)}(0)=\delta_{m_{i}m_{f}}\label{eq:3.1}
\end{equation}
and the scattering probability reduces to
\begin{equation}
P(0,\eta,\beta,\delta,\epsilon)=\left|\sum_{l=0}^{\infty}\sum_{m=-l}^{l}4\epsilon^{2}(l+\frac{1}{2})\frac{(l+|m|)!}{(l-|m|)!}\frac{1}{\Lambda^{2|m|}}e^{-2\epsilon^{2}(\Lambda-bp)^{2}}\mu_{|m|}^{2}(2\epsilon^{2}\Lambda bp)\,e^{i2\sigma_{l}(p)}e^{-(\delta-\xi_{l})^{2}/8}\right|^{2}.\label{eq:3.2}
\end{equation}

\begin{figure}
\begin{centering}
\includegraphics[width=7cm]{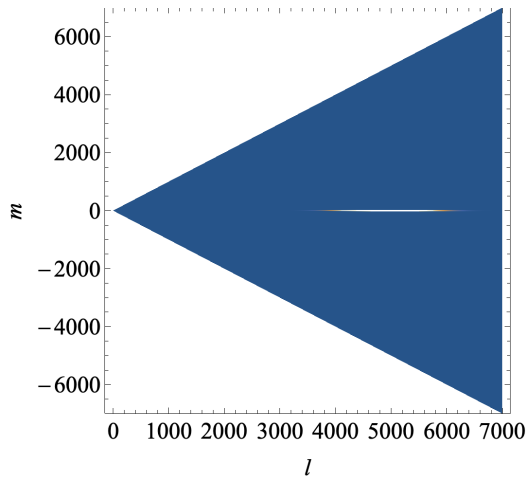}
\par\end{centering}
\caption{\label{fig:Probability-density-in}Probability density in $l$ and
$m$ for $\beta=10$ (absolute values not determined).}

\end{figure}

We can simplify this expression if we only consider impact parameters
$\beta\geq10.$ Then the probability distribution in $l$ is centred
on the magnitude of the classical angular momentum $L=bp,$ with a
width of order $1/\epsilon=1000.$ So low values of $l$ are suppressed.
The classical angular momentum vector has vanishing component in the
$z$ direction, a consequence of our choice of scattering geometry.
So the probability distribution in $m$ is centred on $m=0$ with
a width that we estimate as $\Delta m\sim\beta/\sqrt{2}$ (see Eq.
(\ref{eq:4.6})). These features are shown in Figure \ref{fig:Probability-density-in},
where we plot $|\Phi_{\mathrm{free}}(l,m)|^{2}.$

Then we have
\begin{equation}
\frac{|m|}{l}\sim\frac{\epsilon}{2}\ll1.\label{eq:3.3}
\end{equation}

Then we can use the Stirling approximation (\cite{Gradsteyn1980},
their Eq. (8.327)) for the factorials
\begin{equation}
\frac{(l+|m|)!}{(l-|m|)!}=l^{2|m|}\{1+\mathcal{O}(\frac{|m|}{l})\}.\label{eq:3.4}
\end{equation}
In this approximation $\Lambda$ can be replaced by $l,$ and
\begin{equation}
\Lambda^{2|m|}=l^{2|m|}\{1+\mathcal{O}(\frac{|m|}{l})\}.\label{eq:3.5}
\end{equation}

In Appendix A we derive an asymptotic approximation for $\mu_{|m|}(z),$
for $z\gg1,$ uniform in $m$
\begin{equation}
\mu_{|m|}(z)\sim\frac{e^{-(m^{2}-\frac{1}{4})/2z}}{\sqrt{2\pi z}}.\label{eq:3.6}
\end{equation}
Here $z$ will be of order $z\sim\beta^{2}/2\geq50.$

\begin{figure}
\begin{centering}
\includegraphics[width=7cm]{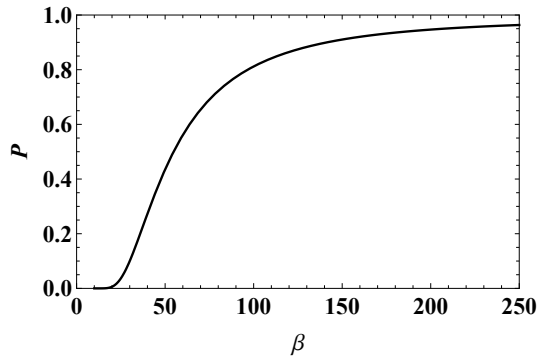}
\par\end{centering}
\caption{\label{fig:Probability-of-scattering}Probability of scattering into
the forward direction as a function of impact parameter for $\eta=22.8.$}

\end{figure}

Then the simplified probability is
\begin{align}
P(0,\eta,\beta,\delta,\epsilon) & \sim\left|\sum_{l=0}^{\infty}4\epsilon^{2}(l+\frac{1}{2})e^{-2\epsilon^{2}(l-bp)^{2}}\sum_{m=-l}^{l}\frac{e^{-(m^{2}-\frac{1}{4})/2\epsilon^{2}lbp}}{4\pi\epsilon^{2}lbp}\,e^{i2\sigma_{l}(p)}e^{-(\delta-\xi_{l})^{2}/8}\right|^{2}\nonumber \\
 & \sim\left|\sum_{l=0}^{\infty}4\epsilon^{2}(l+\frac{1}{2})\frac{e^{-2\epsilon^{2}(l-bp)^{2}}}{\sqrt{8\pi\epsilon^{2}lbp}}\,e^{i2\sigma_{l}(p)}e^{-(\delta-\xi_{l})^{2}/8}\right|^{2}.\label{eq:3.7}
\end{align}
We evaluated this expression numerically for $\eta=22.8,$ on $10\leq\beta\leq250.$
No correction was made for time shifts, which were found to be small
($\delta$ was set to 0). The result is shown in Figure \ref{fig:Probability-of-scattering},
and confirms our hypothesis.

Note that there is another matter to consider. This result is for
alpha particles scattering off a single bare nucleus. In an experiment,
the alphas at high impact parameter and low momentum transfer would
be scattering off effectively neutral atoms, screened by the electrons.
This would make the forward flux even larger than this prediction.

That the forward probability remains small for low impact parameters
is an interesting prediction, but would not be measurable unless the
impact parameter of a collision could be controlled to much less than
an Angstrom.

\section{\label{sec:Behaviour-at-small}Deviations from the Rutherford formula
at low scattering angles}

In a Rutherford scattering experiment described in \cite{Melissinos2003},
the source was the $E=5.2\,\mathrm{MeV}$ emission from Polonium-210
($\eta=21.9$). Those researchers were able to collimate their source
so that the unscattered beam extended to approximately $4^{\circ}=0.070\,\mathrm{rad}.$
We suppose that this could be done with an $E=4.8\,\mathrm{MeV}$
source. From Figure \ref{fig:Shadow-zone-for}, we see deviations
from the Rutherford formula for the zero impact parameter prediction
on $0\leq\theta\leq0.2,$ for $\epsilon=0.001.$ Our estimate for
the size of the deviation region is \cite{Hoffmann2017a}
\begin{equation}
\theta_{D}=4\epsilon|\eta|,\label{eq:4.1}
\end{equation}
proportional to $\epsilon.$ So deviations could possibly be observed
for $\epsilon\geq10^{-3}$ but not for $\epsilon=10^{-4}.$ Note that
the smallest angle (other than the forward direction) at which measurements
were taken in the aforementioned experiment was $6^{\circ}=0.10\,\mathrm{rad}.$

Making predictions just with the $\beta=0$ profile does not give
an accurate portrayal of an experiment. Yet it gives good agreement
with experiment on $\theta\geq0.2.$ An integration is needed over
the distribution of impact parameters, vectors in the $xy$ plane
according to Figure \ref{fig:Scattering-geometry-with}. We label
them with a radial coordinate, $\beta,$ and an azimuthal angle, $\varphi.$

We will implement this integration shortly, but first we need a result
from it to justify a step in the procedure. At $\theta=\pi/2,$ far
from the possible deviation region, we find, after correcting for
a time shift,
\begin{equation}
P(\frac{\pi}{2},22.8,\beta,\varphi,0.6,0.001)\sim P_{\mathrm{Ruth}}(\frac{\pi}{2},22.8,0.001)\,e^{-b^{2}/\sigma_{x}^{2}},\label{eq:4.2}
\end{equation}
independent of $\varphi.$ Then we have
\begin{equation}
\frac{\int d^{2}b\,P(\frac{\pi}{2},22.8,\beta,\varphi,0.6,0.001)}{\int d^{2}b\,e^{-b^{2}/\sigma_{x}^{2}}}=\frac{\int d^{2}b\,P_{\mathrm{Ruth}}(\frac{\pi}{2},22.8,0.001)\,e^{-b^{2}/\sigma_{x}^{2}}}{\pi\sigma_{x}^{2}}=P_{\mathrm{Ruth}}(\frac{\pi}{2},22.8,0.001).\label{eq:4.3}
\end{equation}
With the area $A=\pi\sigma_{x}^{2}$ in the denominator, this averaging
just returns the Rutherford prediction, as required. Then we take
the same area at all angles (and any $\epsilon$) and replace
\begin{equation}
P(\theta,22.8,0,0,\delta,\epsilon)\rightarrow\frac{\int d^{2}b\,P(\theta,22.8,\beta,\varphi,\delta,\epsilon)}{\pi\sigma_{x}^{2}}=\frac{1}{\pi}\int_{0}^{\infty}\beta\,d\beta\int_{0}^{2\pi}d\varphi\,P(\theta,22.8,\beta,\varphi,\delta,\epsilon).\label{eq:4.4}
\end{equation}

The initial and final state vectors are modified to
\begin{align}
|\,\boldsymbol{p}_{i},-R\hat{\boldsymbol{z}}+\boldsymbol{b}(\varphi);\mathrm{in}\,\rangle & =U(R_{z}(\varphi))\,|\,\boldsymbol{p}_{i},-R\hat{\boldsymbol{z}}+b\hat{\boldsymbol{x}};\mathrm{in}\,\rangle,\nonumber \\
|\,\boldsymbol{p}_{f},R_{y}(\theta)\{+R\hat{\boldsymbol{z}}+\boldsymbol{b}(\varphi)\};\mathrm{out}\,\rangle & =U(R_{y}(\theta+\pi))U(R_{z}(-\varphi))A(\mathcal{T})\,|\,\boldsymbol{p}_{i},-R\hat{\boldsymbol{z}}-b\hat{\boldsymbol{x}};\mathrm{in}\,\rangle.\label{eq:4.5}
\end{align}
This introduces a factor $\exp(-i(m_{i}-m_{f})\varphi)$ inside the
sums in Eq. (\ref{eq:2.20}).

\begin{figure}
\begin{centering}
\includegraphics[width=14cm]{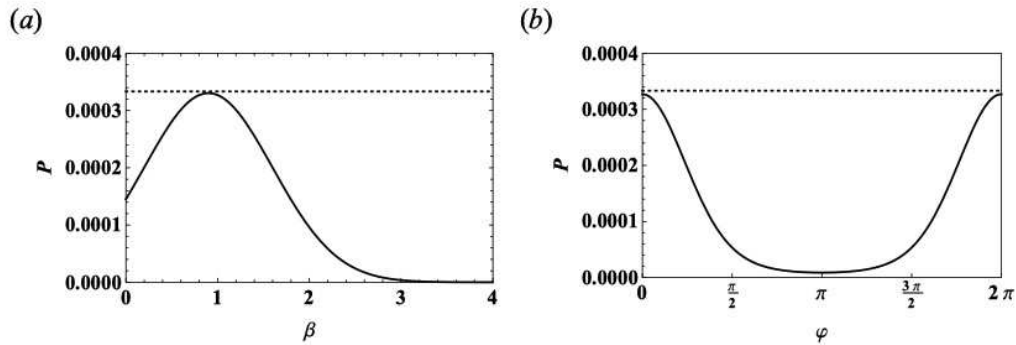}
\par\end{centering}
\caption{\label{fig:Dependence-of-probabilities}Dependence of probabilities
on (a) $\beta$ and (b) $\varphi$ at $\theta=0.1.$}

\end{figure}

We will only consider scattering angles in $0.1\leq\theta\leq0.2.$
Then we will find that it suffices to integrate up to a scaled impact
parameter $\beta=3$ (see Figure \ref{fig:Dependence-of-probabilities}
(a)). The width in $m$ of the function $\mu_{|m|}(2\epsilon^{2}\Lambda bp)$
is, from Eq.~(\ref{eq:3.6}),
\begin{equation}
\Delta m=\sqrt{2\epsilon^{2}\Lambda bp}\sim\sqrt{2\epsilon^{2}(bp)^{2}}=\frac{1}{\sqrt{2}}\beta,\label{eq:4.6}
\end{equation}
using $\Lambda\sim bp$ at the wavefunction peak. So we expect that
it will be sufficient to consider $m_{f},m_{i}$ values in Eq. (\ref{eq:2.20})
in the range $|m_{f}|,|m_{i}|\leq2.$

Again we are able to approximate
\begin{equation}
[\frac{(l+|m|)!}{(l-|m|)!}]^{\frac{1}{2}}\frac{1}{\Lambda^{|m|}}\sim1\label{eq:4.7}
\end{equation}
and replace $\Lambda$ with $l.$

Since we are only considering small angles, we can use the small angle
approximation of the Wigner rotation matrices found in \cite{Hoffmann2018b}.
There are problems with evaluating Wigner rotation matrices of large
order in M{\small{}ATHEMATICA} \cite{Mathematica2019}. These problems
are avoided with this approximation in terms of Bessel functions.
We use
\begin{equation}
d_{m_{i}m_{f}}^{l}(\theta)\sim(-)^{m_{i}-m_{f}}(\frac{\theta}{\sin\theta})^{\frac{1}{2}}J_{m_{i}-m_{f}}(l\theta)\quad\mathrm{for}\ m_{i}\geq m_{f}\label{eq:4.8}
\end{equation}
and
\begin{equation}
d_{m_{1}m_{2}}^{j}(\theta)=(-)^{m_{1}-m_{2}}d_{m_{2}m_{1}}^{j}(\theta).\label{eq:4.9}
\end{equation}

Summing the relevant terms in Eq. (\ref{eq:2.20}) gives
\begin{align}
P(\theta,\eta,\beta,\delta) & \sim\frac{\theta}{\sin\theta}|\sum_{l=0}^{\infty}4\epsilon^{2}(l+\frac{1}{2})\,e^{-2\epsilon^{2}(l-bp)^{2}}e^{i2\sigma_{l}(p)}e^{-(\delta-\xi_{l})^{2}/8}\{J_{0}(\mu_{0}^{2}+2\mu_{1}^{2}+2\mu_{2}^{2})-iJ_{1}\mu_{2}\mu_{1}4\cos\varphi-iJ_{1}\mu_{1}\mu_{0}4\cos\varphi\nonumber \\
 & \quad-J_{2}\mu_{2}\mu_{0}4\cos(2\varphi)-J_{2}\mu_{1}^{2}2\cos(2\varphi)+iJ_{3}\mu_{2}\mu_{1}4\cos(3\varphi)+J_{4}\mu_{2}^{2}2\cos(4\varphi)\}|^{2},\label{eq:4.10}
\end{align}
where
\begin{equation}
J_{n}=J_{n}(l\theta),\quad\mu_{n}=\mu_{n}(\epsilon l\beta),\quad n=0,1,2,\dots.\label{eq:4.11}
\end{equation}

Using this, we found the probability at $\theta=\pi/2$ to be independent
of $\varphi,$ as discussed above. At $\theta=0.1,$ for example,
we find dependence on both $\beta$ and $\varphi,$ as seen in Figure
\ref{fig:Dependence-of-probabilities}.

\begin{figure}
\begin{centering}
\includegraphics[width=7cm]{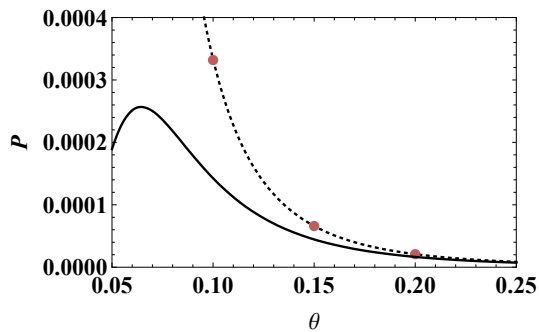}
\par\end{centering}
\caption{\label{fig:Results-of-integration}Results of integration over impact
parameter, showing probabilities as points. The Rutherford probability
(dotted) and the zero impact parameter prediction (solid) are shown
for comparison}

\end{figure}

Evaluating Eq. (\ref{eq:4.10}) numerically for $\theta=0.10,0.15,0.20$
gives the results shown in Figure \ref{fig:Results-of-integration}.
The Rutherford probability and the zero impact parameter prediction
are shown for comparison. Remarkably, the integration over impact
parameters supplies the missing probability needed to give excellent
agreement with the Rutherford formula.

We have not done an analysis of the propagation of errors from our
approximation procedures. We note that the differences of our predictions
from the Rutherford formula for $\theta=0.10,0.15,0.20$ are all less
than $1\,\%.$

This process cannot continue to arbitrarily small scattering angles.
The Rutherford ``probability'', Eq. (\ref{eq:2.25}), reaches unity
at
\begin{equation}
\theta_{1}=\epsilon\sqrt{8|\eta|}=0.014\label{eq:4.12}
\end{equation}
with our parameters, so deviations are inevitable below that angle.
For impact parameters with $\beta\geq10,$ the character of the probability
profile changes, as we will discuss in a future paper. The major contributions
to probability then come at angles smaller than the minimum considered
here, $\theta=0.1.$ Deviations from the Rutherford formula are then
probable, after integration over impact parameters. Of course measurements
at low angles would require an incoming beam collimated to a small
angular width. This would decrease the count rate, but in this region
the expected count rates are the highest.

A significant unknown in this analysis is the size of the wavepackets
from a given source, or the distribution of sizes. If a source was
producing wavepackets with $\epsilon\leq2.1\times10^{-4},$ as was
discussed in Section \ref{sec:Scattering-into-the}, there would be
little chance of ever measuring deviations from the Rutherford formula.
However it may be possible to \textit{increase} the fractional momentum
width, $\epsilon,$ from a source, say by increasing the thickness
of the radioactive layer. The value $\epsilon=0.001$ considered here
was chosen as low but not excessively low, with the error factor $\sqrt{\epsilon}$
acceptably low. Unfortunately, the current state of alpha particle
spectrometers \cite{Ruddy2006} is not sufficient to resolve a linewidth
with $\Delta E/E=2\epsilon=0.002.$

\section{\label{sec:Conclusions}Conclusions}

The aim of this paper was to investigate whether the shadow zone of
low scattering probability around the forward direction predicted
for zero impact parameter would persist when nonzero impact parameters
were considered. The first calculation we performed indicated that,
for events with impact parameters much larger than the wavepacket
width, the wavepackets would exit into the forward direction largely
undisturbed by the interaction. This confirms an obvious feature of
a Coulomb scattering experiment: the strong signal in the forward
direction from the unscattered part of the beam.

Then we found that events with impact parameters of order the wavepacket
width introduce significant probability into the region of deviation
from the Rutherford formula (for $\beta=0$). For a beam with a distribution
of impact parameters, averaging over impact parameters is necessary
to construct a prediction. We did this for three low angles in the
region of deviation and found, remarkably, predictions in excellent
agreement with the Rutherford formula. This extends the validity of
that formula further down into the deviation zone.

There may still be deviations from that formula that could be measured
experimentally, with a well-collimated alpha beam (at the expense
of lower count rates). Further investigation, focussing on even larger
impact parameters, will be done in a future paper.

\appendix

\section{Asymptotic approximation of $\mu_{|m|}(z)$}

The defining differential equation for the modified Bessel functions,
$I_{\nu}(z),$ is (\cite{DLMF2020}, their Eq. (10.25.1))
\begin{equation}
\{\frac{d^{2}}{dz^{2}}+\frac{1}{z}\frac{d}{dz}-(1+\frac{\nu^{2}}{z^{2}})\}I_{\nu}(z)=0.\label{eq:a.1}
\end{equation}
The previously known asymptotic approximation for large $z$ is (\cite{DLMF2020},
their Eq. (10.40.1))
\begin{equation}
I_{\nu}(z)\sim\frac{e^{z}}{\sqrt{2\pi z}}\{1-\frac{\nu^{2}-1/4}{2z}+\mathcal{O}(\frac{1}{z^{2}})\}.\label{eq:a.2}
\end{equation}
This does not capture the dependence on $\nu$ for large $\nu,$ so
we seek an asymptotic approximation uniform in $\nu$ that holds for
large $\nu.$

We write
\begin{equation}
I_{\nu}(z)=\frac{e^{x}}{\sqrt{2\pi x}}g_{\nu}(z).\label{eq:a.3}
\end{equation}
Then we find the differential equation for $g_{\nu}(z)$ is
\begin{equation}
g_{\nu}^{\prime\prime}+2g_{\nu}^{\prime}-\frac{\nu^{2}-1/4}{z^{2}}g_{\nu}=0.\label{eq:a.4}
\end{equation}
For large $z$ we scale this equation, writing
\begin{equation}
z=M\xi,\label{eq:a.5}
\end{equation}
with $M\gg1$ and $\xi$ of order unity. This gives
\begin{equation}
\frac{1}{M^{2}}\frac{d^{2}g_{\nu}}{d\xi^{2}}+\frac{2}{M}\frac{dg_{\nu}}{d\xi}-\frac{1}{M^{2}}\frac{\nu^{2}-1/4}{\xi^{2}}g_{\nu}=0.\label{eq:a.6}
\end{equation}
Note that for $\nu$ of order $\sqrt{M},$ the third term becomes
of the same order as the second.

Ignoring the first term gives
\begin{equation}
\frac{dg_{\nu}}{dz}\cong\frac{\nu^{2}-1/4}{2z^{2}}g_{\nu},\label{eq:a.7}
\end{equation}
with solution
\begin{equation}
g_{\nu}(z)=C\,e^{-(\nu^{2}-\frac{1}{4})/2z}.\label{eq:a.8}
\end{equation}
We set $C=1$ to find agreement with Eq. (\ref{eq:a.2}) for small
$\nu.$ From our definition of $\mu_{|m|}(z),$ Eq. (\ref{eq:3.6}),
we have the asymptotic approximation
\begin{equation}
\mu_{|m|}(z)\sim\frac{e^{-(m^{2}-\frac{1}{4})/2z}}{\sqrt{2\pi z}}.\label{eq:a.9}
\end{equation}

We use this result to approximate the sum
\begin{align*}
S(z) & =\sum_{m=-l}^{l}\mu_{|m|}^{2}(z)\\
 & \sim\sum_{m=-\infty}^{\infty}\frac{e^{-(m^{2}-\frac{1}{4})/z}}{2\pi z}\\
 & \sim\frac{1}{\sqrt{4\pi z}}.
\end{align*}
A numerical check showed a fractional error in this approximation
less than 0.0013 for $z\geq50.$

\bibliographystyle{vancouver}

\end{document}